# Multimodal Imaging System Combining Hyperspectral and Laser Speckle Imaging for In Vivo Hemodynamic and Metabolic Monitoring

Junda Wang[a], Luca Giannoni*[a], Ayse Gertrude Yenicelik[b],

Eleni Giama[b], Frédéric Lange[a], Kenneth J. Smith[b]

and Ilias Tachtsidis[a]

[a]Department of Medical Physics and Biomedical Engineering, University College London, London, UK

[b]Department of Neuroinflammation, UCL Queen Square Institute of Neurology, London, UK

*Corresponding author: Luca Giannoni; l.giannoni@ucl.ac.uk
Other authors e-mails: zczqj43@ucl.ac.uk; ag.yenicelik@ucl.ac.uk; eleni.giama.23@ucl.ac.uk; f.lange@ucl.ac.uk; k.smith@ucl.ac.uk; i.tachtsidis@ucl.ac.uk

**Abstract** We present the development and validation of a novel multimodal optical imaging platform that integrates hyperspectral imaging (HSI) and laser speckle contrast imaging (LSCI) to enable real-time, non-invasive mapping of tissue oxygenation, perfusion and metabolism, via blood flowmetry and targeting of oxy- ($HbO_2$), deoxyhemoglobin (HHb), as well as oxidized cytochrome-c-oxidase (oxCCO). The system's architecture features a single high-speed camera and dual optical path, with synchronized alternating illumination: a filtered, supercontinuum laser for HSI and a He-Ne laser for LSCI. The system's performances were evaluated through *in vivo* experiments on rat spinal cord under normoxic and hypoxic conditions, revealing coherent physiological changes in hemodynamics, metabolism and relative blood flow index (rBFI). These results demonstrate the potential of the platform for functional tissue imaging and quantitative dynamic monitoring of both oxygen delivery and consumption.

**Introduction** HSI and LSCI are complementary optical techniques for monitoring tissue physiology. HSI provides detailed spectral information at each pixel, enabling quantification of tissue biochemical properties such as $HbO_2$, HHb, and oxCCO – markers of oxygenation and metabolic state [1]. In contrast, LSCI measures dynamic speckle patterns to map blood flow (perfusion) in real time, with high temporal resolution but without direct chemical specificity. Individually, each modality has limitations: HSI typically has lower frame rates due to wavelength scanning and cannot directly assess perfusion, while LSCI yields only relative flow information and lacks metabolic insight. By combining HSI and LSCI, a more comprehensive picture of tissue health can be obtained, capturing both metabolic status and hemodynamic changes simultaneously. Previous studies have demonstrated the benefit of multimodal imaging in contexts like surgical guidance and wound monitoring [2][3]. However, most prior multimodal implementations relied on parallel architecture (two cameras and distinct optical paths), necessitating complex alignment procedures to merge data. These setups can suffer from parallax, field-of-view differences, and timing mismatches between modalities [3][4].

In this work, we developed a novel, integrated HSI-LSCI imaging system that uses a single camera and a single optical path to overcome the limitations of previous approaches. Our system acquires hyperspectral and speckle images in rapid alteration using synchronized illumination sources, effectively achieving quasi-simultaneous, dual-modal imaging on the same field of view. This sequential strategy ensures inherent spatial co-registration (each pixel corresponds to the same tissue region in both modalities) and obviates the need for post hoc image alignment or dual-camera calibration. The innovation lies in the coordinated hardware and software design that enables millisecond-scale switching between broadband spectral illumination and laser speckle illumination, with precise timing control to capture dynamic physiological events in real time. By integrating HSI's biochemical specificity with LSCI's fast perfusion mapping, the system can monitor oxygenation and blood flow simultaneously, providing insight into tissue oxygen supply-demand balance and vascular responses that would be difficult to attain with either modality alone.

The primary application of the developed imaging system is to study hemodynamic and metabolic coupling *in vivo*. We focused on the rat spinal cord as a model, where blood flow and oxygen metabolism are critical, for example, in injury or ischemia. The spinal cord experiments involved altering the inspired oxygen levels of the subject to induce physiological changes. This allowed us to test the system's capability to detect concurrent changes in tissue oxygenation and perfusion under both normal and hypoxic conditions.

**Materials and Methods** The hardware configuration (illustrated schematically in Fig.1) is an upgrade of the hNIR system developed by Giannoni *et al.* [3]. The new setup employs a high-performance scientific CMOS camera (Andor Zyla 5.5, 5.5 megapixels, 16-bit dynamic range), ensuring high sensitivity for both HSI and LSCI, with pixel-level co-registration. Spectral illumination for HSI is provided by a supercontinuum la-ser (NKT Photonics FIR20), paired with a motorized Pellin–Broca prism and optical fibers to generate 11 discrete, narrow (8-11 nm, FWHM) bands (600, 630, 665, 784, 800, 818, 835, 851, 868, 881, and 894 nm). Each HSI frame is acquired with exposure times of 100–200 ms. LSCI is performed using a coherent He-Ne laser (632.8 nm), expanded to illuminate the field uniformly, and captured with milli-second-scale exposures. A custom optical assembly with a 15× reflective objective and infinity-corrected tube lens delivers <10μm spatial resolution for both modalities. A PC-controlled acquisition system synchronizes camera triggering, wavelength tuning, and laser modulation.

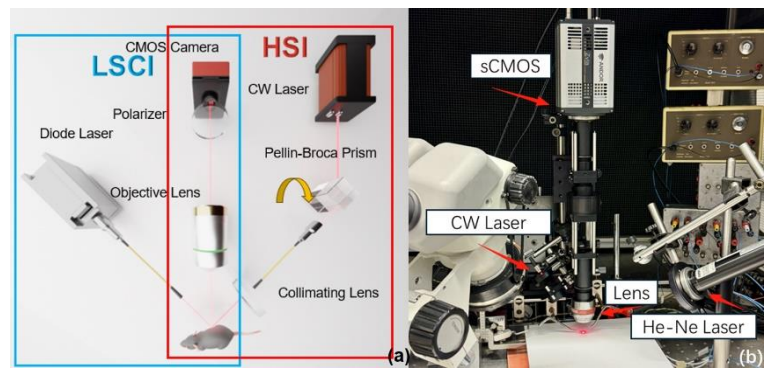

***Fig. 1*** *(a) Schematic diagram of the integrated HSI-LSCI imaging system, showing the single optical path shared by the hyperspectral and speckle subsystems. (b) Photograph of the assembled system, highlighting the camera, broadband light source with tunable filter (Pellin–Broca prism), laser illumination optics, objective lens, and the sample platform.*

To evaluate the system's performance, *in vivo* imaging experiments were conducted on rats, focusing on the spinal cord. All animal procedures wereperformed under appropriate anaesthesia and with ethical approval. An incision and dorsal laminectomy (removal of part of vertebra) at the T12–L2 level were carried out to expose the thoracolumbar spinal cord for imaging. The rat was placed prone on a temperature-controlled stage to maintain body temperature, and the exposed spinal cord was kept moist with sterile saline. The imaging system was positioned above the spinal cord such that the region of interest on the cord was centred in the camera's field of view. To minimize motion, the animal was secured, and breathing was maintained at a steady rate under anaesthesia. An opaque enclosure shielded the area from external light to prevent spectral contamination during HSI.

The rat was subjected to a controlled sequence of respiratory gas conditions to induce changes in tissue oxygenation. First, a baseline period under normal air (21% $O_2$, normoxia) was recorded for 5 minutes. Next, the inspired gas was switched to a hypoxic mixture (10% $O_2$ in nitrogen) for 3–5 minutes to create moderate systemic hypoxia. After the hypoxic phase, the gas was returned to normal air (21% $O_2$) to allow recovery, and imaging was continued for another 5 minutes to observe reperfusion or hyperemic responses as oxygen levels normalized. Finally, an extreme hypoxia phase was introduced after recovery by further lowering the inspired oxygen to 5% $O_2$ for a few minutes (in these cases, imaging was stopped when the animal showed signs of agonal breathingand was euthanized per ethical guidelines). The imaging continued without interruption, capturing the transition from normoxia to hypoxia and the reverse transition back to normoxia. Local tissue oxygenation and temperature were measured using an implantable optical probe (OxyMicro, WPI, USA) inserted into the spinal cord throughout all experimental phases, to ensure that physiological data were interpreted in the context of the animal's condition.

The data analysis for *in vivo* studies involved computing quantitative maps from the recorded HSI and LSCI frames. Hyperspectral data were processed to derive tissue concentration changes of $HbO_2$, HHb, and oxCCO using Modified Beer-Lambert's law applied on each pixel [3]. Speckle images were converted to rBFI maps by calculating the speckle contrast (the ratio of the standard deviation to mean intensity in a small kernel) and relating it to flow speed. For each experiment, regions of interest on the spinal cord were selected, and average values of the concentration changes of $HbO_2$, HHb, oxCCO and rBFI in these ROIs were tracked over time and compared against the tissue oxygen saturation provided by the physiological monitoring.

**Results** Using the described protocol, we obtained simultaneous blood flow, oxygenation and metabolism from the rat spinal cord during transitions from normoxia to hypoxia (and back). Fig. 2 presents representative pseudo-color maps of hemodynamic parameters captured in the spinal cord. In normoxia (21% $O_2$), the maps of $HbO_2$ concentration, HHb, and rBFI all showed stable, baseline values with minimal spatial variation – indicating well-perfused, well-oxygenated tissue at rest. Upon switching to 10% $O_2$ (moderate hypoxia), significant physiological changes were observed. Oxygenation dropped, and the concentration of $HbO_2$ in the spinal cord decreased by roughly 16 μm·cm (pathlength-normalized units) from baseline reflecting increased extraction of oxygen by the tissue as the supply diminished. The oxidation state of oxCCO also fell by about 6 μm·cm in the hypoxic period. The drop in oxCCO suggests that cellular aerobic metabolism was moderately impaired due to reduced oxygen delivery. These spectral changes were not uniform across the field: the pseudo-color HSI maps revealed heterogeneous regions where oxygenation decreased more markedly, highlighting that some areas of the spinal cord became more hypoxic than others [5].

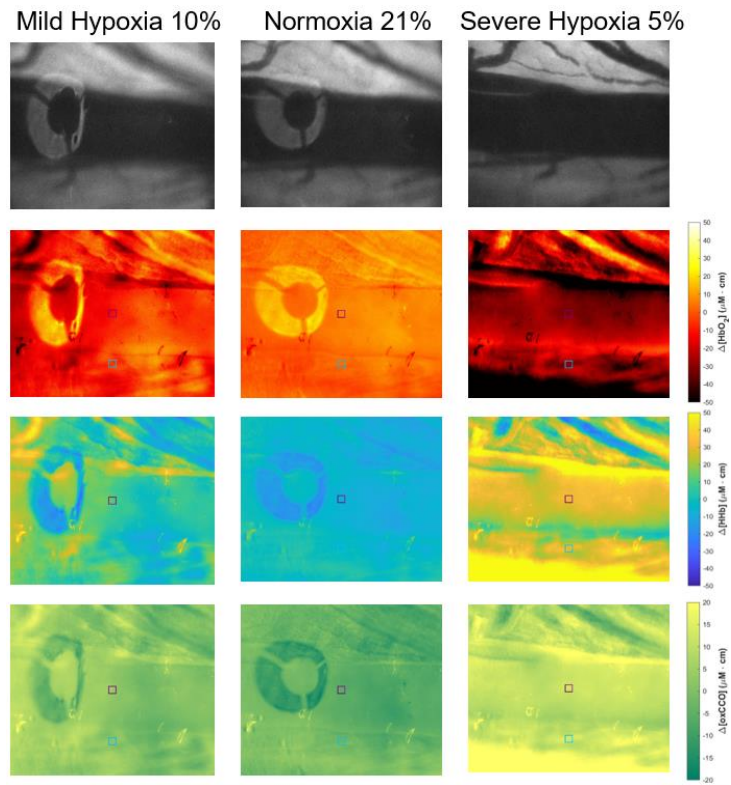

***Fig. 2*** *Raw image after histogram equalization adjustment (8-bit), and HSI experimental results compare with baseline. The purple and blue boxes indicate the ROI in vessels and tissues. A ring artefact is visible in the image, caused by internal reflection within the reflective objective. This artefact is spatially consistent and does not interfere with interpretation of the central region of interest.*

Simultaneously, the LSCI-derived blood flow signals changed: the relative blood flow index maps showed that perfusion was affected in tandem with oxygenation. At 10% $O_2$, certain regions of the spinal cord displayed reduced rBFI values, indicating decreased blood flow in those areas, and these regions corresponded spatially to those with the largest drops in oxygenation. In other words, areas that became most deoxygenated were also experiencing low perfusion, underscoring the close coupling between blood supply and tissue oxygen levels. When severe hypoxia (5% $O_2$) was applied in the subset of experiments, the spinal cord's response intensified. As expected, oxygenation metrics showed further decline: $HbO_2$ concentrations fell additional 10–11 μm·cm compared to normoxia (though interestingly the incremental drop from the mild to severe hypoxia was smaller than the drop from normoxia to mild hypoxia). This smaller change suggests that by the time the tissue reached 10% $O_2$, it may have already extracted as much oxygen as possible (approaching physiological extraction limits. HHb levels in severe hypoxia remained high (about 10–11μm·cm above baseline, similar to the mild hypoxia increment), confirming that blood in the spinal cord was highly deoxygenated. The oxCCO dropped further (another 3–4 μm·cm) in severe hypoxia, indicating significant mitochondrial oxygen starvation and likely reduced ATP production.

Correspondingly, the blood flow maps in Fig. 3 showed widespread reduction in perfusion; by this stage, the circulation itself may be failing (in extreme hypoxia leading to bradycardia or circulatory collapse, perfusion in the microvasculature can dramatically decrease). The spatial maps at 5% $O_2$ clearly illustrate large areas with very low $HbO_2$ and low rBFI, aligning with each other, which is consistent with the notion that without blood flow, oxygen delivery ceases and tissue oxygenation plummets. Throughout these experiments, the concurrent imaging of HSI and LSCI provided insight into the temporal dynamics of the physiological responses.

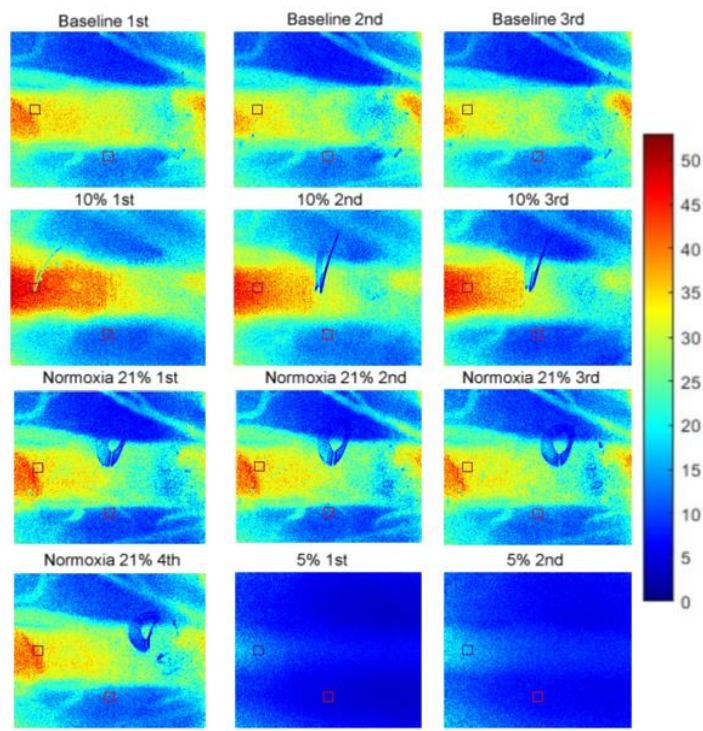

***Fig 3.*** *rBFI at baseline, 10% oxygen, return to normoxia, and 5% oxygen. The scale shows rBFI (inverse of speckle contrast). The purple and red boxes indicate the ROIs.*

Fig. 4 presents representative time-series of $HbO_2$, HHb, oxCCO, and rBFI during a normoxia–hypoxia–recovery protocol. Under baseline normoxic conditions, $HbO_2$ and oxCCO remained stable, while HHb levels were low, consistent with well-oxygenated and metabolically active tissue. Upon introduction of 10% $O_2$, $HbO_2$ and oxCCO declined markedly, indicating reduced oxygen delivery and impaired mitochondrial oxidative metabolism. HHb simultaneously increased due to enhanced oxygen extraction. Notably, rBFI exhibited a transient rise immediately after hypoxia onset, likely reflecting compensatory vasodilation or autoregulatory mechanisms, before decreasing to a new, lower plateau. During reoxygenation, $HbO_2$, oxCCO, and rBFI all rebounded—often with overshoots—suggesting reactive hyperemia and restored metabolic function, while HHb returned toward baseline. In experiments involving 5% $O_2$, the decline in $HbO_2$ and oxCCO was more abrupt and profound, with rBFI collapsing, consistent with severe circulatory and metabolic failure. These results illustrate the tight coupling between hemodynamic and metabolic parameters under hypoxic stress and recovery.

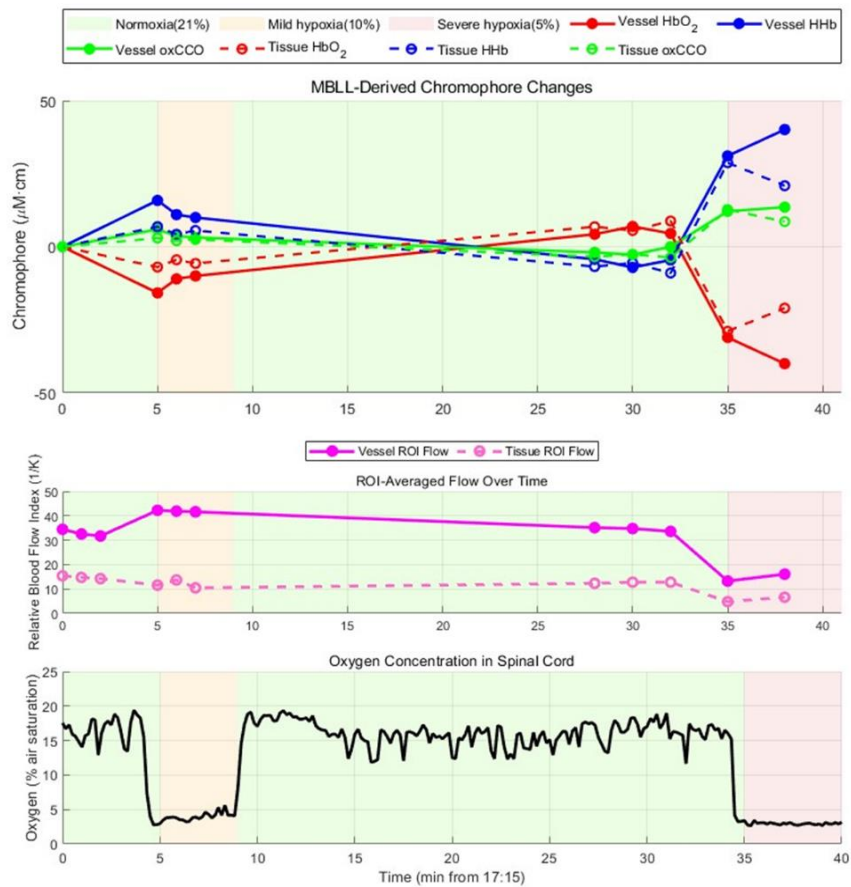

***Fig. 4*** *Temporal plots of spinal cord oxygenation and perfusion under graded hypoxia. (Top) HSI-derived changes in chromophore concentrations (ΔHbO₂, ΔHHb, ΔoxCCO) within a selected region of interest (100×100 pixels), across normoxia (21% O₂, green), mild hypoxia (10% O₂, orange), and severe hypoxia (5% O₂, red) phases. (Middle) rBFI derived from LSCI within the same region. (Bottom) Measured oxygen concentration in the spinal cord expressed as % air saturation*

The spinal cord experiments under varying oxygen conditions provided a clear demonstration of the system's utility – revealing how the spinal cord's blood flow and metabolism react together to oxygen lack. The ability to detect these changes simultaneously and in real time is what distinguishes this work. The data underscore important physiological relationships (flow-metabolism coupling) and validate the approach for studies of neurovascular health.

**Discussion and Conclusions** This study introduced a combined HSI-LSCI imaging system and demonstrated its performance *in vivo*. The major findings and contributions are the following:

- We successfully developed a single-platform system that monitors blood flow and oxygenation together, providing a more comprehensive assessment of tissue status than single-modality imaging. The co- registered measurements enabled us to directly correlate perfusion with oxygenation on a pixel level, which is invaluable for identifying phenomena like localized ischemia or reactive hyperemia. The results clearly show the advantage of the multi-modal approach. For example, certain pathological changes (such as impending tissue damage) might only be evident when looking at both perfusion and metabolic data in tandem.

- By applying controlled oxygen level changes, we observed distinct physiological responses captured by the system. Under mild hypoxia (10% O₂), the spinal cord exhibited increased oxygen extraction (drop in HbO₂, rise in HHb) accompanied by perfusion adjustments (evidence of vasodilation to sustain flow). Under severe hypoxia, oxygenation levels fell dramatically and could not be maintained despite any perfusion compensation, and eventually blood flow also declined, indicating failure of autoregulation and onset of tissue damage. These findings mirror known compensatory limits and illustrated that our system can track the dynamic interplay between oxygen supply and utilization in real time.

- A significant technical achievement of this work is the real-time fusion of hyperspectral and speckle imaging data. We were able to process and display oxygenation and blood flow maps concurrently during experiments, with minimal latency. To our knowledge, this is one of the first implementations of truly integrated, on-line HSI-LSCI for *in vivo* studies.[6]

The system's temporal resolution is limited by sequential acquisition, with each HIS-LSCI cycle running at ~1-2 Hz, potentially missing rapid events. Future improvements include snapshot HSI or faster cameras to boost speed. LSCI provides only relative superficial flow, requiring calibration. Motion artifacts remain a challenge, suggesting motion correction or respiratory gating. Hardware upgrades—such as brighter, uniform illumination and more sensitive cameras—could improve image quality and accuracy. Overall, this technology provides a powerful tool for biomedical research, allowing investigators to observe how changes in perfusion and metabolism coincide, in real time and at high spatial resolution.[7] With further development, such integrated imaging approaches could translate to clinical applications, where they might be used for monitoring tissue viability during surgeries, assessing stroke or spinal cord injury patients, or guiding therapies that target blood flow and oxygen delivery. The ability to see both the supply (blood flow) and the outcome (tissue oxygenation) at once offers a more complete understanding of tissue health than either measure alone – a step toward more informed diagnostics and interventions in healthcare.